\documentstyle[12pt]{article}

\newfont{\sechead}{cmr10 scaled\magstep1}
\begin{document}

\title{WHERE IS TOP?}

\vskip0.20in

\author{R.H. Dalitz*, Gary R. Goldstein$\dagger$}\vskip0.30in
\date{\null}
\maketitle
\vskip0.10in
\noindent
*Department of Physics, University of Oxford, Theoretical Physics,
1
Keble Road
, Oxford OX1 3NP, U.K.\\
$\dagger$ Department of Physics, Tufts University, Medford,
Massachusetts
02155, USA\vskip0.30in

\underline {ABSTRACT}
\vskip0.20in
\noindent Possibilities are discussed for determining the top quark
mass $m_t$ from observations on the decay processes for top-antitop
pairs produced in antiproton-proton collisions at Tevatron
energies, assuming that the $t \to bW^+$ decay channel is dominant
and much faster than hadronization. The final states $t \bar{t} \to
\bar{b} b \mu^\pm  e^\mp$ provide the most striking signal, with
little background, but they are rare ($\approx 2/81$). If all three
candidate events prove to be from $t \bar{t}$, an estimate follows
for $P(m_t \mid rate)$, the probability distribution for $m_t$. The
one reported configuration allows an independent estimate for
$P(m_t \mid \mu^\pm e^\mp \,2jets)$. These two distributions are
compatible and can be combined to give an $m_t$ estimate of about
122 GeV. Decay events ``1 energetic lepton($l$) + 4jets'' should
appear twelve times as often as ``$\mu^\pm e^\mp \,2jets$'' events and
can be analysed to give estimates for $P(m_t \mid l\, 4jets)$; this
is illustrated for a fictitious event. There may be background from
non-top events but suitable cuts on the data and our analysis
procedure together reduce this to a low level. The rate observed
for these events does not appear to be as large as this factor 12.
Identification of either or both of the $(b \bar{b})$ jets would be
a great step forward, separating out the ``$\mu^\pm e^\mp b
\bar{b}$'' and ``$l b \bar{b} \,2jets$'' events of importance with
negligible background. We advocate an energetic approach to the
analysis of individual events (whether 2,1 or no ($b$ or $\bar{b}$)
jets identified) on an event by event basis, with the hope of
finding a subgroup of events with a common mass estimate.\footnote{This
paper was given by the first author (R.H.D.) on 19 July 1992
at the International School of Subnuclear Physics at ERICE. Its
text has been updated to the end of 1992.}

\newpage

\section{\underline {{\sechead INTRODUCTION}}}

\subsection{\sechead Existence of the top quark}
\vskip -0.05in
\hspace{0.2in}~~~~ There is little doubt today about the
existence of a top quark
$t$, the partner to the well-known and much-studied bottom quark
$b$.
The Standard Model, with SU(2) x U(1) for the Electroweak
Interactions
has had remarkable success, many of its parameters now being known
to
high accuracy.  In particular, the weak-isospin component
$T^{wk}_3$
for
the b quark has now been determined empirically in a very direct
way,
from measurements of the forward-backward asymmetry in the
process $e^+e^- \to b \bar{b}$ and of the decay
width $\Gamma(Z^0 \to b \bar{b})$.  From these data, Kane and Kolda
[1]
deduced the following $T_3^{wk}$ values for the b-quarks.
$$
 T_3^{wk}(b_L) = -0.49^{+0.05}_{-0.02}
\hspace{0.05in},\hspace{0.2in}
T^{wk}_3(b_R) = 0.00^{+0.10}_{-0.08}\hspace{0.05in}, \hspace{ 2in}
(1.1)
$$

\noindent which are in accord with the Standard Model assignments of the
b quark to the third quark-lepton family, for which the leptonic member
$\tau^+$ is already
known, as shown in Table 1.
\vskip 0.2in

Table 1. The three families in the SU(2) X U(1) Standard Model
\vskip 0.1in
\begin{tabular}{|cc|ccc|} \hline & & & & \\
$T^{wk}_3$ & $Y^{wk}$ & \multicolumn{3}{c|}{Quark and Lepton states
for each family} \\ & & & & \\ \hline
$+\frac{1}{2}$&$+\frac{1}{3}$&$u_{\scriptscriptstyle
L}$&$c_{\scriptscriptstyle
L}$&$(t_{\scriptscriptstyle L})$ \\ & & & & \\
$-\frac{1}{2}$&$+\frac{1}{3}$&$d_{\scriptscriptstyle
L}$&$s_{\scriptscriptstyle
L}$&$b_{\scriptscriptstyle L}$ \\ & & & & \\
0&$\{+\frac{4}{3}$,$-\frac{2}{3}\}$&$\{u_{\scriptscriptstyle
R}$,$d_{\scriptscriptstyle R}\}$&$\{c_{\scriptscriptstyle
R}$,$s_{\scriptscriptstyle R}\}$&$\{(t_{\scriptscriptstyle
R})$,$b_{\scriptscriptstyle R}\}$ \\ & & & & \\
$+\frac{1}{2}$&$+1$&$\bar{\nu}_{\scriptscriptstyle
eL}$&$\bar{\nu}_{\scriptscriptstyle \mu
L}$&$(\bar{\nu}_{\scriptscriptstyle \tau L})$ \\ & & & & \\
$-\frac{1}{2}$&$+1$&$e_{\scriptscriptstyle
L}$&${\mu}_{\scriptscriptstyle
L}$&${\tau}_{\scriptscriptstyle L}$ \\ & & & & \\
0&$+2$&$e_{\scriptscriptstyle R}$&${\mu}_{\scriptscriptstyle
R}$&${\tau}_{\scriptscriptstyle R}$ \\  \hline
\end{tabular}
\vskip 0.2in
\noindent The fact that the
predictions of the (spontaneously broken) SU(2) x U(1) symmetry
have
been so successful numerically provides a powerful argument that
the
third (quark-lepton) family must be complete.  The absence of a
$T^{wk}_3 = + {1 \over{2}}$ partner to the $b_L$ quark would be
such a
gross violation of this symmetry that it would no longer be
possible
to understand the detailed and widespread agreement of the data
with
the predictions based on this electroweak symmetry.  In short,
there
must exist a $T^{wk}_3 = + {1 \over{2}}$ partner to the bottom
quark
$b_L$, and it has been natural for this partner to be named the top
quark t.

{}~~~~Empirically, little else is known about the top quark, although
its interactions are prescribed in form and magnitude from its
place in
the third quark-lepton family.  However, we do know that its mass
$m_t$
is very much greater than the mass of the b quark, $m_b \approx
5~GeV$.
A firm limit on $m_t$ is provided by the fact that top-antitop
pair creation is not observed in $e^+e^-$ annihilation in the
energy region of $Z^0$ excitation; in other words, the
threshold for $t\bar{t}$ production must be above $M_Z$,
giving the lower limit\\
$$ m_t ~\stackrel{>}{\sim} M_Z/2 \approx 46 ~GeV. $$

{}~~~~Indeed, the top mass is now believed to be above 91 GeV [2].
If top were lighter than this, ~$t\bar{t}$ pair production would
have been observed already at the Tevatron,
where proton-antiproton interactions are studied
at c.m. energy 1800 GeV, an energy far more than adequate
for any reasonable value for $m_t$.  At the Tevatron, the
limiting factor for top mass determination is the rate of
events, rather than the available energy.  We shall discuss
this further in Sec.2.

{}~~~~The purpose of current top quark research is not just to
demonstrate
 directly the existence of the top quark (which we believe already)
nor
to check that its interactions are in accord with SU(2) x U(1)
symmetry,
 although the latter studies will provide a very fruitful field for
research later on.  The top quark will not lose its polarization by
hadronization, as do the other heavy quarks, charm and
bottom(c.f. Sec.1.2),
 so that it will be possible to test all the detailed spin
dependences of its
interactions [3-5].  It may even turn out to be possible to use the
top quark
decay as a spin analyser for heavy particle and/or high
energy processes which
happen to give rise to top quarks.  No, the paramount purpose of
present top research is to determine the mass $m_t$ of the top
quark; how this
measurement may be achieved will be discussed in Sec.3 below.  Only
after
this mass has been determined can we move on with the Standard
Model, e.g.
with the determination of the Higgs particle mass and
with
the testing of more extensive symmetry models which contain SU(2)
x U(1) as a
subgroup.  Only then will we know what kind of accelerator we
shall need for all
these detailed studies.
\vskip 0.20in
\subsection{\sechead The top quark lifetime }
\vskip -0.05in
\hspace{0.2in}~~~~When $m_t$ exceeds $(M_W + m_b) \approx 85 GeV$, top quark
physics becomes qualitatively different from the physics we have
become
accustomed to for the c and b quarks, as was first pointed out by
Bigi [6].
 The dominant decay mode for the top quark is then the two-body
mode
$$
(a)~~~ t \rightarrow W^++b, \hspace{0.5in} (b)~~~ \bar{t}
\rightarrow
W^- + \bar{b} \hspace {2.5in} (1.2)
$$

\noindent As the mass $m_t$ increases, this decay process becomes
faster than hadronization.  The top decay lifetime calculated
for this mode [7] is plotted as function of $m_t$ on Fig.1.
For $m_t$ appreciably lower than ($M_W + m_b$), the W-boson is
virtual and its leptonic decay mode generates the overall mode
$t \to l^+ \nu_lb$, whose partial lifetime falls like $m_t^{-5}$.
In the transition region, as $m_t$ approaches and passes ($M_W +
m_b$),
the decay lifetime falls even more rapidly with $m_t$.  For $m_t$
well above this threshold, as holds for the physical
situation, the decay lifetime falls more slowly, ultimately
like $m_t^{-3}$.  The calculated partial width is given as
function of $m_t$ in Table 2.
\vskip0.1in

\begin{center}
{}~Table 2.  Partial decay width for $t \to bW^+$, as function of
$m_t$[29].
\vskip0.1in
\begin{tabular}{|c|cccccc|} \hline
$m_t(GeV)$  & 100 & 120 & 140 & 160 & 180 & 200 \\ \hline
$\Gamma(t \to bW^+)(MeV)$ & 88 & 298 & 612 & 1033 & 1572 & 2242 \\
\hline
\end{tabular}
\end{center}
\vskip 0.1in

{}~~~~[By way of contrast, we comment briefly on the other
heavy quarks $Q =c$ and $b$.  Their decay lifetimes are of order
$10^{-12}$ sec. In $c$ and $b$
jets, a
polarized quark $Q$ forms a meson $(Q\bar{q})$ where $q$ denotes a
light quark
 $q = (u, d, s)$, by picking up a light antiquark $\bar{q}$ from
the vacuum.
  If this meson is in the $^1S_0$ ground state, all polarization
carried by $Q$ is lost, since this state is
spherically symmetric and
cannot carry spin information.  If the meson state $(Q\bar{q})$
which is
formed has non-zero spin, e.g. the $^3S_1$ state or
orbitally-excited
states like $^3P_2$, it undergoes fast hadronic or
electromagnetic transitions (since these conserve parity, the
angular distributions do not depend on any initial polarization,
although they can transfer polarization from the initial to the
final state) until it reaches the $^1S_0$ state, which is the
lowest state for the $(Q\bar{q})$ system.  Thus, all of the
polarization information carried initially by the quark $Q$ is
lost,
for the cases $Q=(c,b)$[8].]
\vskip0.05in

\section{\underline{{\sechead TOP MASS ESTIMATES FROM EMPIRICAL
DATA}}}

\subsection{\sechead Virtual top: radiative corrections}
\vskip -0.05in

\hspace{0.2in}~~~~It was first pointed out by Veltman [9] that,
although the
$e^+e^-$ energy at LEP is too low for the production of real
top-antitop
pairs, the existence of the top quark can still have a substantial
effect on the predictions of the Standard Model through the
 ``radiative corrections'' generated by virtual top-quark
loops and exchanges.  Since the Standard Model is a renormalizable
theory, these corrections are computable, at least in a
perturbative approximation.  Now that LEP experiments have measured
with high accuracy many quantities related with the electroweak
interactions, these measurements can be compared with the
corrected theoretical predictions in order to draw some
conclusions concerning the top quark and any other particles of
high mass.  Some of these measurements are the masses
$M_Z$ and $M_W$, the total width $\Gamma_Z$ and some partial
widths for $Z^0$ decay, and the forward-backward asymmetries
for $e^+e^- \to \bar{b}b$ and $e^+e^- \to$ leptons
in an energy range covering the $Z^0$ peak, which arise
from $\gamma$-$Z^0$ interference.
In the minimal Standard Model, there are several other
parameters also relevant,
 namely $M_H$, the Higgs particle mass, and $\alpha_S(M^2_Z)$, the
QCD
coupling strength evaluated at the $Z^0$ mass.
The latter can be deduced with fair accuracy from purely hadronic
processes
as well as from electroweak studies in the $ Z^0 $ mass range.
Given the recent LEP data and the theoretically computed
expressions, it is then possible to lay out on an
$(m_t, M_H)$ plane, the regions consistent with these data.
Quite a number of analyses have been carried out along
these lines recently [10,11].  With the LEP data updated
to July 1992, Ellis et al. [10] have given the value\\
$$
all~data \mid_{M_{H}~free}: ~~~~~~ m_t = 124 (27) GeV,
\hspace{2.5in} (2.1)
$$\\
using $\alpha_S(M^2_Z) = 0.118(8)$.
Their analysis has some sensitivity to $M_H$; adopting
by choice a rather low value for $M_H$, their analysis gives\\
$$
M_H = M_Z : m_t = 132 (25) GeV. \hspace{3.2in} (2.2)
$$\\
requiring a small increase for the optimum $m_t$.

{}~~~~More elaborate models generally have more free parameters.
The Minimal Supersymmetric Standard Model (MSSM) has aroused much
interest recently, owing to its success in extrapolating the
three coupling strengths, $\alpha_i(\mu^2)$, at scale $\mu^2$ =
 (momentum transfer)$^2$ and with $i = (1,2,3)$ appropriate to the
strong and electroweak sectors of the SU(3) x SU(2) x U(1) symmetry
contained within a Minimal SU(5)SUSY, back to a common value at a
GUT scale $\mu$ of order $10^{16}~ GeV$ [12], a test that an
earlier attempt based on SU(5) $\supset$ SU(3) x SU(2) x U(1)
had failed [13].  Besides $M_H$ and $m_t$, this MSSM introduces
five new parameters, $m_A$ and tan$\beta$ for the Higgs sector,
$(m _{\stackrel{\sim}{g}}, m_0)$ the gluino mass and a common
sfermion mass, respectively, and $\Lambda_t$, a
supersymmetry-breaking parameter.  The discussion of the
allowed regions for all of these parameters is naturally
rather complicated.  Ellis {\it et al.\/} [14] have discussed these
dependences only for $m_t = 130~GeV$, mainly to illustrate
how an empirical value for $m_t$ would give much-needed
information on the bounds which constrain the MSSM parameters.
Most probably, given the number of new parameters introduced,
the limits placed on $m_t$ in the MSSM will be much less
restrictive than those for the minimal SM without supersymmetry.

\subsection{{\sechead Total cross section for top-antitop
production}}
\vskip -0.05in

\hspace{0.2in}~~~~At the Tevatron energy of 900~GeV for proton
and for antiproton, the (900+900)~GeV $\bar{p}p$ interactions
are rather like (300+300)~GeV interactions between a valence
quark q and a valence antiquark $\bar{q}$.  Creating a
$t \bar{t}$ pair, each with (say) $m_t = 150~GeV$, through
the interaction process\\
$$
 \bar{q} + q \to \bar{t} + t \hspace{4.1in} (2.3)
$$\\
absorbs half of the initial energy, not leaving much energy for the
creation of more particles.  The energies of the residual $(qq)$
and $(\bar{q}\bar{q})$ quarks from the proton and antiproton,
respectively, lead through hadronization to particles which
mostly go down the beam pipe.  Thus, after the production process,\\
$$ \bar{p} + p \to \bar{t} + t + X, \hspace{3.65in} (2.4)$$\\
the $t$ and $\bar{t}$ quarks decay according to (1.4 a,b), giving\\
$$\bar{p} + p \to \bar{b} + W^- + b + W^+ + X, \hspace{2.65in}
(2.5)$$\\
where X consists of (i) $X_{inv}$ consisting of hadrons which
go down the beam pipe, and (ii) $X_{vis}$, the other hadrons
recorded by the detector.  The $W^\pm$ bosons then decay,
with lifetime 3.1 x 10$^{-25}s$.  Their simplest and most
visible decay processes are\\
$$(a) ~~ W^+ \to l^+ + \nu_l~, \hspace{0.05in} (b) ~~ W^- \to l^-
+
\bar{\nu}_l~, \hspace{2in} (2.6)$$\\
for the three leptons $ l = e, \mu$ and $\tau $.

{}~~~~The most striking events are those where the $W^+$ and $W^-$
decays lead to two charged leptons from different families without
any
$ X_{vis} $, a typical final state being\\
$$ \bar{b} + b + \mu^\pm + e^\mp + (\nu_\mu + \bar{\nu}_e +
X_{inv}),\hspace{2.4in} (2.7)$$\\
where the $\bar{b}$ and $b$ quarks hadronize to give corresponding
jets $j(\bar{b})$ and $j(b)$.  There may also be some secondary
jets, emitted from the initial quarks or by the quarks heading
for the beam-pipe, or arising from the development of the $b$
and $\bar{b}$ jets, but these will generally have relatively
low energy, so that these $(\mu^\pm e^\mp b \bar{b})$ final
states may be expected to have rather simple structure.
Another attractive feature of these $(\mu^\pm e^\mp b \bar{b})$
events is that they may be expected to be rather free of
background,
since $\mu$ and $e$ belong to different families.  There is
no particle $Y$ known whose decay leads to the $\bar{\mu}e$
or $\mu \bar{e}$ configuration without corresponding neutrinos
$\nu_e (or~ \bar{\nu}_e)$ and $\bar{\nu}_\mu (or~ \nu_\mu)$.
Indeed, lepton conservation is believed to hold separately
for muonic and electronic lepton numbers: strong limits have been
placed
on any violation of this conservation law, through searches for
reactions of the type $\mu^- p \to e^- p$ and for decays of the
type $Y \to \mu^\pm e^\mp X$, especially for $Y^0 = K^0_L$ and $Y^+
= K^+$.

{}~~~~ The simplest $\bar{p}p$ process leading to final
states with $(\mu^\pm e^\mp)$ is $W$-pair production\\
$$ \bar{p} + p \to W^- + W^+ + n ~jets, \hspace{2.8in} (2.8)$$\\
followed by the decays (2.6).  This process involves two weak
vertices
 whereas the process (2.4) has none; provided that the energy
available
is reasonably large compared with $2(M_W + m_b)$, the process (2.8)
with $n=2$ may be expected to have a rate smaller by two
factors of $(\alpha_{em}/2 sin^2\theta_W)$ than that for (2.5),
if the $n=2$ jets were $b$-jets.  However, the process (2.8)
will be in fact dominated by events with light quark jets in
large numbers, and it is difficult to estimate how much
background (2.8) will produce for the $\bar{t}t$ events (2.4)
leading to final state (2.5).  It is a question of what
fraction of these non-$t\bar{t}$ events (2.8) having $n=2$
light quarks and surviving the transverse momentum cuts
($cf$. Secs. 3 and 4) can be adequately fitted to the
interpretation (2.7).  The only further remark we can make
is that, if a microvertex detector is used, so that it is
possible to add the requirement that two of the jets in (2.8)
be identified as $b$ jets, the background rate from (2.8)
would be of order $\{\alpha_{em}/(2 \alpha_s sin^2\theta_W)\}^2
\approx 10^{-2}$ times the signal for (2.7), following the
top-pair production (2.4).  In our further discussion here
and in Sec. 3, we shall neglect background to the process (2.5)
arising from the direct production of $W$ pairs.  However, the
validity of this neglect is under discussion at present.

{}~~~~ The cross section for $\bar{t}t$ production in $\bar{p}p$
annihilations
at the Tevatron energy of 900~GeV through reaction (2.4) has been
calculated
by Eichten [15], as function of $m_t$.  At Tevatron energies and
for
such
large $m_t$ values ( $~\stackrel{>}{\sim}$ 120 GeV) as we are led
to
consider here,  $\bar{t}t$ production is dominantly due to the
quark-antiquark process (2.3); production by gluon-gluon processes
is small in comparison.

{}~~~~ The $W^\pm$ bosons have hadronic modes, in addition to the
leptonic
modes (2.6).  These are due to the $W$-coupling with the quark
weak-isospin charged current.  Since we are interested
here only in total rate, we limit our discussion to the couplings\\
$$ W^+ \to \hspace{0.09in} (a)\hspace{0.09in} u + \bar{d} ,
\hspace{0.3in} (b) \hspace{0.09in} c + \bar{s},\hspace{2.5in}
(2.9)$$\\
given by diagonal terms of the Cabibbo-Kobayashi-Maskawa matrix.
Each
of these couplings leads to two final quark jets. The third
coupling
is $W^+ (t + \bar{b})$ whose threshold lies far above $M_W$, so
that it
 contributes only to virtual processes, such as the radiative
corrections mentioned above.

{}~~~~ Recognizing colour, there are 6 couplings in all,
each with the same strength as each $W^+ \to l^+\nu_l$.
Empirically, the branching ratio for each $(l \nu_l)$
final state is 10.5(9)\%, which agrees with the naive
expectation of 1/9, from the channels just enumerated.  If
both $W^+$ and $W^-$ in (2.5) decay hadronically, the
final state will have 6 energetic quark jets, a configuration
which it may be difficult to disentangle.
If only one $W$ decays hadronically, the final state will
generally have one energetic lepton and four quark jets.  The
analysis of this final state is generally possible and we shall
discuss this in Sec 3.  The dilepton events have been described
in part just above.  The decays (2.6) include the possibility
of $\tau^\pm$ emission; this is far more difficult to deal with
experimentally than is $e^\pm$ or $\mu^\pm$.  For this reason,
we shall not refer explicitly to the possibility of $\tau^\pm$
emission again.  On the other hand, final electrons or muons can
be easily recognized and distinguished.  Among the dilepton final
states, the $(e^+e^-)$ and $(\mu^+\mu^-)$ cases need special
attention and much caution, since there are many other mechanisms
which can give rise to electron or muon pairs, one example
being the production and decay of a $Z^0$ boson, and which
will still require special cuts for their exclusion.  In fact,
no candidate event of this type has yet been reported, although
such events ($e^+e^-~$and$~\mu^+\mu^-$) must have the same
rate as the $(\mu^\pm e^\mp)$ events.

\subsection{\sechead Top mass estimate from production rate}
\vskip -0.05in

\hspace{0.2in}~~~~ In  this section, we confine attention to
the $(\mu^\pm e^\mp)$
events.  The expected number $N_{\mu e}$ of these events is plotted
as
function of $m_t$ in Fig.2 for an integrated
luminosity $IL = 30 p b^{-1}$, based on the estimates by
Crane [16] who used Eichten's calculation of the cross section
for top-antitop pair production [15] and took due account of the
efficiency $\kappa$ for the CDF detector as a function of the
event location and configuration.
For mass $m_t$, we have\\
$$ N_{\mu e} = \kappa \cdot IL \cdot \sigma_{\mu e}(m_t)\>, \hspace{4.2in}
(2.10)$$\\
where $\sigma_{\mu e}$ denotes the cross section for $(\mu^\pm
e^\mp)$
events in $\bar{p}p$ collisions at 1800~GeV c.m. energy.  The
probability of producing $n$ such events, given $m_t$, is\\
$$ P(n \mid m_t) = \frac{N_{\mu e}(m_t)^n}{n!} exp (-N_{\mu
e}(m_t))\hspace{3.1in}
(2.11)$$\\
Given $IL,~ \sigma_{\mu e}(m_t)$ and the observation
of $n$ events, the Bayesian probability distribution for $m_t$ is
then\\
$$P(m_t \mid n) = \frac{N_{\mu e}(m_t)^n}{n!} exp (-N_{\mu
e}(m_t))/\{\int dm \frac{N_{\mu e}(m)^n}{n!} exp (-N_{\mu
e}(m)\},\hspace{0.8in} (2.12)$$\\
where the integral is taken over all possible values $m$
allowed for the top quark $t$ by all the conservation laws.
This probability may be reduced to the form\\
$$P(m_t \mid n) = c(n)(\kappa\cdot IL\cdot \sigma_{\mu e}(m_t))^n.
exp(-\kappa \cdot
IL \cdot \sigma_{\mu e}(m_t)),\hspace{1.93in} (2.13)$$\\
$c(n)$ being an $n$-dependent normalization factor.
This distribution peaks at the $m_t$ value for which\\
$$\sigma_{\mu e}(m_t) = n/\kappa \cdot IL\>.\hspace{4.4in}(2.14)$$

{}~~~~ One good $(\mu^- e^+)$ candidate event has already been
published by
the CDF collaboration [2] and has been discussed [17] in some
detail
($cf.$ Sec. 3 below).  A second $(\mu e)$ candidate was shown by
the
CDF collaboration in their report given at the November 1992
Chicago
Meeting
of the Division of Particles and Fields of the American Physical
Society,
although no measurement details were released.  It was well known
at
that
meeting that the DO collaboration also had their first $(\mu e)$
candidate.
\footnote{This event  was shown at a Fermilab seminar in February
1993, but
without full details.}  Although the integrated luminosities IL are
not
known to us precisely, a value of about 20 $pb^{-1}$ for CDF
(including IL=4.7 $pb^{-1}$ from their 1989 paper) and 10 $pb^{-1}$
for
DO would appear plausible estimates, at least of the right order
of magnitude. The mean detector efficiency can be deduced from
a comparison of Crane's rates with Eichten's total $t\bar{t}$
production
cross sections.

{}~~~~ On the assumption that these three $(\mu e)$ candidates do
stem
from top-antitop production, and that the integrated luminosity up
to November 1992 was about 30 $pb^{-1}$, the probability
distribution
for $m_t$ is shown on Fig. 3.   Its peak is at 120 GeV, the
one-deviation limits being 109 and 135 GeV.  Since the curve
for $N_{ \mu e}(m_t)$ shown in Fig.2 is a steeply falling
function of $m_t$, the peak value thus determined for $m_t$
is not strongly dependent on our estimate for $IL$, nor on the
number of $\mu e$ events.  For $(IL, n) = (40 pb^{-1}, 3)$, the
peak value is $m_t = 127.5$; for (30$pb^{-1}, 4)$, it is at 114
GeV,
and for $(30 pb^{-1},2)$, it is at 129 GeV.

\section{\underline{{\sechead ANALYSIS OF DILEPTON EVENTS}}}
\subsection{\sechead Kinematics of top decay sequence $ t
\rightarrow b W^{+} \rightarrow b {\overline{l}}^{+} {\nu}_l $ }
\vskip -0.05in

\hspace{0.2in}~~~~ Consider first the kinematics of top
decay to $bW^+$, followed by $W^+$ decay to
$\bar{l}\nu_l$, in any frame.  Energy-momentum conservation gives\\
$$ t = b + \bar{l} + {\nu}_{l}.\hspace{4.7in} (3.1)$$\\
where $a$ denotes the energy-momentum four-vector of the
particle a.  Since $\bar{l}$ and $\nu_l$ are decay products of $W$,
we have\\
$$M^2_W = (\bar{l} + \nu_l)^2 = (t - b)^2,\hspace{3.75in} (3.2)$$\\
and, since the neutrino has zero mass,\\
\begin{tabular}{lllr}
0& = &$(t - b - \bar{l})^2$&\hspace{3.05in}$(3.3a)$\\
& = &$(t-b)^2 - 2 \bar{l}.(t + b) +
\bar{l}^2.$&$\hspace{3.05in}(3.3b)$
\end{tabular}

\noindent where $a.b$ denotes the scalar product of
the four-vectors $a$ and $b$ and $a^2 = a.a$.
For $l = e$ or $\mu$, we can neglect the
lepton mass, i.e. $l^2 = m_l^2 = 0$.
Using (3.2) in (3.3b), we obtain the result\\
$$ \bar{l}.t = \bar{l}.b + M^2_W/2.\hspace{4.32in} (3.4)$$\\
Evaluating the r.h.s. in the lab. frame and l.h.s. in
the top rest-frame, we deduce that\\
$$E_{lt} = (\bar{l}.b + M_W^2/2)/m_t,\hspace{3.8in} (3.5)$$\\
giving $E_{lt}$, the lepton energy in the
top rest-frame, in terms of lab. measurements for $\bar{l}$ and
$b$.

{}~~~~It is useful to derive here an inequality for $m_t$.  From
(3.1)
and (3.2), we have\\

\begin{tabular}{lllr}
$m^2_t$ & = & $(b + \bar{l} + {\nu}_{l})^2$ &$\hspace{2.95in}
(3.6a)$\\
& = & $m^2_b + M^2_W + 2b.\bar{l} + 2b.{\nu}_{l}$ &$
\hspace{2.95in}(3.6b)$\\
\end{tabular}\\
Evaluate the invariant products $b.\bar{l} $~and~$b.{ \nu}_{l}$ in
the $W$ rest frame, where $E_l = l$ and $E_\nu = \nu$ are each
$M_W/2$,
and the momenta $\underline{\bar{l}}$ and $\underline{{\nu}_{l}}$
are
opposite.  Denoting by $\theta$ the angle between
$\underline{\bar{l}}$ and $\underline{b}$ in this frame,
their product takes the form\\

\begin{tabular}{lllr}
$b.\bar{l}~ b.{\nu}_{l}$ & = & $(E_b - bcos\theta)(E_b + bcos
\theta
)(M_W/2)^2 $&$\hspace{1.95in} (3.7a)$\\
& = & $(m^2_b + b^2 sin^2 \theta)(M^2_W/4) $&$
\hspace{1.95in}(3.7b)$\\
& $\geq$ & $m^2_bM^2_W/4. $&$\hspace{1.95in} (3.7c)$\\
\end{tabular}

\noindent Using this inequality for b.${\nu}_{l}$ in Eq.(3.6b),
leads to the inequality\\

\begin{tabular}{lllr}

$m^2_t$ & $\geq$ & $m^2_b + M_W^2 + 2b.\bar{l} +
m^2_bM^2_W/2b.\bar{l}
$&$ \hspace{2.45in}(3.8a)$\\
& = & $(m^2_b + 2b.\bar{l})(M^2_W + 2b.\bar{l})/2b.\bar{l}$&$
\hspace{2.45in}(3.8b) $\\

\end{tabular}

\noindent as given in the Appendix of ref.[5].

\subsection{\sechead A geometric construction}
\vskip -0.05in

\hspace{0.2in}~~~~Now consider the kinematics in the lab. frame,
starting with
the 3-momenta $\underline{b}$ and $\underline{\bar{l}}$.
The top momentum $\underline{t}$ is constrained by two relations,\\
$$ (\underline{t}-\underline{b})^2 = (E - E_b)^2 - M_W^2,
\hspace{3.4in}
(3.9) $$\\
from eq.(3.2) and\\
$$ (\underline{t}-\underline{b}-\underline{\bar{l}})^2 = (E - E_b
-
E_l)^2, \hspace{3.2in} (3.10) $$\\
from eq.(3.3a), where E has not been constrained to the value
$E_t$ because the value of $m_t$ is still left open.  For
given $E$, eqs. (3.9) and (3.10) constrain $\underline{t}$
to lie on the intersection of two spheres, one centred on the
point $\underline{b}$ and the other centred on
$(\underline{b} + \underline{\bar{l}})$.  Their intersection is a
circle on a plane perpendicular to the line BL joining these two
centres and centred on this line.  As $E$ varies, the centre of
this circle moves along the line $BL$ at a rate linear in $E$,
and the square of its radius also increases linearly with $E$;
in short, as $E$ varies, this circle traces out a paraboloid
with axis $BL$, as described in refs. [8,17].  All points
$\underline{t}$ satisfying the constraints (3.2) and (3.3a) for
given $\underline{b}$ and $\underline{\bar{l}}$, for all possible
$m_t$ values, lie on this paraboloid.  The points of interest
to us are those for a definite $m_t$, still to be determined.
These points lie on an ellipse formed by the intersection of the
paraboloid by a plane whose normal lies in the plane $OBL$ and
makes
an angle $\sigma$ with the axis $BL$, where\\
$$ tan \sigma = b_1/(E_b - b_3),\hspace{3.85in} (3.11)$$
is independent of $m_t$, and $b_1$ and $b_3$ are defined on
Fig.4.  For a specified $m_t$, the possible vectors
$\underline{t}$ are given by $\vec{OT}$, as $T$ moves around this
ellipse, and the vector $\vec{LT}$ gives the corresponding
neutrino momentum $\underline{{\nu}_{l}}$.  It is apparent
that the top quark energy $E_t = \surd(m_t^2 + \underline{t}^2)$
must lie between two limits $E_{min}(m_t) \leq E_t(m_t) \leq
E_{max}(m_t)$.

{}~~~~As $m_t$ increases, the plane PTQ moves upwards, with a
constant normal since the angle $\sigma$ (given by (3.11)) does not
depend on $m_t$.  The ellipse retains the same eccentricity but
increases in linear dimension.  As $m_t$ decreases, the plane moves
downward and the ellipse shrinks until for a limiting value $m_t =
m_*$ the plane becomes tangent to the paraboloid and the ellipse is
reduced to a point.  For $m_t < m_*$, there is no solution for
$\underline{t}$.  This value $m_*$ is the lowest limit for an $m_t$
consistent with the momenta $(\underline{b} ,\underline{\bar{l}})$;
$m_*$ is, of course, the value given by the square root of the
expression (3.8b).  The projection of the ellipse for $m_t$ on a
plane perpendicular to the vector $\underline{\bar{l}}$ is a circle,
and the centre of the ellipse projects onto the centre of the
circle. As $m_t$ varies, the centre of the ellipse moves parallel
to the vector $BL = \underline{\bar{l}}$ and is distant from it by
amount $(M_W^2 /2E_l) tan\sigma$ in the plane of $\underline{b}$
and $\underline{\bar{l}}$.

{}~~~~It is worth noting here that any configuration $(\underline{b}
,\underline{\bar{l}})$ can be fitted by a sufficiently large $m_t$.  However, a
very large $m_t$ means that LT is very large, but this is the neutrino momentum
$\underline{\nu_l}$, of which we have no direct knowledge.  It means also that
OT must be correspondingly large, and this is the top momentum $\underline{t}$.
These two very large momenta would be compensating each other, differing by the
observed momentum $(\underline{b} +\underline{\bar{l}})$, and this is not
generally a plausible interpretation of the top decay event observed.

{}~~~~For top-antitop production, with lepton emission in both decays, two
paraboloids are to be constructed, one from $(\underline{b}
,\underline{\bar{l}})$ for t and one from $(\underline{\bar{b}} ,
\underline{l})$
for $\bar{t}$, both in the same laboratory frame.  One dilepton event has
already
been described in detail in the literature [17] from the CDF experiment in 1988
as a possible t production event.  We shall refer to it as CDF-~1.  Its
measurements are given here in Table 3, the z-axis being along the beam.  There
is a third lepton in CDF-1, a $\mu^+$ with quite high energy but with rather
small transverse energy.  It allows interpretation as a secondary muon, emitted
from $\bar{B}$ meson decay in the $\bar{b}$-jet following the $\bar{t} \to
\bar{b} W^-$ decay of the hypothetical antitop quark produced in this
$p\bar{p}$
interaction.  It has the right charge sign and a low transverse energy
compatible
with the sequence $\bar{b} \to \bar{c} W^+$, $W^+ \to \mu^+ \nu_{\mu}$, and it
travels in the same general direction as an energetic hadronic jet.  These
features strongly suggest that the $\mu^+$ and the hadronic jet are to be taken
together, as the components of a $\bar{b}$ jet.  We expect that
the other jet is a b quark jet but there is no evidence to demonstrate this.

\subsection{\sechead  Top mass estimation from dileptonic decays
$ \bar{t} \to \mu^\pm e^\mp+2jets$ }
\vskip -0.05in

\hspace{0.2in}~~~~If we now \underline{assume that CDF-1 represents
top-antitop production},
with the assignments given in Table 1, the two paraboloids then constructed are
those shown in Fig.5, projected onto the (y,z) plane. The ellipses representing
t decay are clearly seen;  the $\bar{t}$-ellipses happen to be seen edge-on in
this projection, since the momentum vectors $\underline{\bar{b}}$ and
$\underline{l}$ almost lie in this plane, their components $b_x$ and $l_x$
being
small. These two sets of ellipses are shown for $m_t$ ranging from 115 to 305
GeV, in 10 GeV intervals.  The quark-antiquark interaction which gave rise to
this event was strongly asymmetric, since the two paraboloids lie almost
entirely
within the same hemisphere.

{}~~~~It had been assumed generally, we think (but see ref.[19]), that it would
not
be possible to analyse such an event in terms of top-antitop pair creation
since the final state would then include at least two energetic neutrinos,
$\nu_e$ and
$\bar{\nu}_{\mu}$. [CDF-1 actually has a third neutrino, of type $\nu_\mu$  and
with unknown energy, from the  decay mentioned above. It is assumed here that
this $\nu_\mu$ energy is small although its associated $\mu^+$ has quite high
energy.  This assumption needs checking for internal consistency].  Of course,
we do not know whether this $ \bar{t}$ assumption for event CDF-1 is correct,
but if it is, then it is possible to deduce something probabilistic about
$m_t$.

\begin{center}
{}~Table 3. Input data for two $ \bar{t}$ candidates, one real (CDF-1) and the
other fictitious ($F_{\mu e}$), both of the form ($\mu^\pm e^\mp b \bar{b}$).
\vskip 0.2in
\begin{tabular}{|c|cccc|cccc|} \hline
 & \multicolumn{4}{c|}{Event CDF-1} & \multicolumn{4}{c|}{Event $F_{\mu
e}$} \\ \hline
GeV/c & $e^{+}$ & jet($\bar{b}$) & $e^{-}$ & jet($b$)* & $e^{-}$ &
jet($\bar{b}$)
& $\mu^{+}$ &  jet($b$) \\ \hline
$p_x$ & -21.2 & 18.7 &  0.6 & -0.4  &  22.5 & 66.3 &  8.0 & -47.7 \\
$p_y$ & 23.6 & -6.25 & -43.7 & 16.8 &  -5.45 & 86.4 & 20.0 & 9.5 \\
$p_z$ & -28.55 & 25.25 & 16.8 & -98.9 & -21.5 & -12.0 & 94.0 & 27.5 \\
 & \multicolumn{4}{c|}{*includes a low energy $\mu^{+}$} & \\ \hline
\end{tabular}
\end{center}
\vskip 0.1in
{}~~~~There is one further input.  In the simplest parton picture of the $p
\bar{p}$ collision, [20], each of the partons $q$ and $\bar{q}$ has no
transverse
component of momentum.  It follows that the net transverse momentum of the
final
$t$ and $\bar{t}$ quarks must be zero, so that we have the two further
constraints:\\
$$\underline{\bar{t}}_T + \underline{t}_T = 0.\hspace{3.9in} (3.12) $$\\
where the suffix T denotes the component of the vector transverse
to the incident
proton beam. What is the consequence of this condition?  We can see this most
simply by rotating the  paraboloid by $180^{\circ}$ about the z-axis.  If we
view
the two resulting paraboloids along the z-axis, i.e. project them on to the XY
plane, the constraint equations (3.12) for an assumed $m_t$ correspond to a
crossing of the two ellipses for this $m_t$, since (3.12) then reads\\
$$-\underline{\bar{t}}_T = \underline{t}_T. \hspace{3.9in} (3.13) $$\\
These crossings are shown for a number of $m_t$ values on Fig.6, for the event
CDF-1.  For $m_t < 110.2~GeV$, the ellipses do not cross (for $m_t < 100~GeV$,
they do not even exist) nor do they cross for $m_t > 410~GeV$ (which is an
unreasonably large $m_t$, as is explained below).  For $m_t$ between these
limits, there are two or four crossing points, each giving a solution for
$\underline{t}$ and $\underline{\bar{t}}$ (and hence for $\underline{\nu}_e$
and $\underline{\bar{\nu}}_{\mu}$ separately).  The next step is to assign a
probability to each configuration.

{}~~~~From the z-components of a solution for the momenta ($\underline{t},
\underline{\bar{t}}$), the momenta $xP$ and $\bar{x}P$ of the parton and
antiparton, where P denotes the total momentum of $p$ and $\bar{p}$ in the Lab.
frame, can be deduced, giving\\
$$x = (E_t - E_{\bar{t}} + (t_L + \bar{t}_L))/2P, \hspace{3in} (3.14a)$$\\
and\\
$$\bar{x} = (E_t - E_{\bar{t}} - (t_L + \bar{t}_L))/2P, \hspace{3in}
(3.14b)$$\\
where the suffix L denotes the component of the vector along the incident
proton
beam.  The structure functions for the proton and the antiproton are the same
and
are well-known [21]. For the proton there are three of these, $F(x)$ for the
quark partons, $\bar{F}(x)$ for the antiquark partons and $F_g(x)$ for the
gluons, and they depend on (momentum transfer)$^2$, for which we have taken the
value $(m_t)^2$.  For the antiproton, the functions $F(x) and \bar{F}(x)$
are to
be interchanged and $x$ replaced by $\bar{x}$.  For the values of
($\underline{t},\underline{\bar{t}}$) relevant to the Tevatron experiments, the
relevant values $x$ and $\bar{x}$ are quite large, being  typically 0.1 or
larger.
For such values of $x$ and $\bar{x}$, $\bar{F}$ is small and $F_g$  is even
smaller. We should add here a note that, some way below the upper limit allowed
for $m_t$ by the kinematics,
the values of $x$ and $\bar{x}$ required by eqs.(3.14)
may exceed unity, as $m_t$ increases; of course, the expressions (3.14) are
then
not physically meaningful solutions. Thus, the condition that $x$ and $\bar{x}$
must be less than unity imposes an upper limit on the magnitude of $m_t$.

{}~~~~Thus the $\bar{t}t$ production rate at the Tevatron is dominated by the
quark-antiquark collisions, which are proportional to $F(x) F(\bar{x})$. We
have
used for F set 2 of the Duke and Owens [22] structure functions for (u+d),
neglecting $\bar{F}$ and $F_g$.  Even F(x) falls to zero quite rapidly, like
$(1-x)^3$, as x increases above 0.5.   The rate is also proportional to the
differential cross section for $parton-antiparton \to t\bar{t}$, summed over
all
of the parton-antiparton collisions. Including for completeness the $i =
\{1,2,3\} = \{(\bar{q}q),(q\bar{q}),(gg)\}$ initial states, we find for the
rate
factor,\\
$$R(x,\bar{x}) = \sum_i F_i(x) F_i(\bar{x})
\frac{d\sigma(\hat{s},\hat{t})_i}{d\hat{t}}, \hspace{2.5in} (3.15) $$\\
where $F_1(x)$ now stands for $F(x)$, $F_2(x)$ for $F(\bar{x})$ and $F_3(x)$
for
$F_g(x)$, and $\hat{s} = x\bar{x}s$, and $\hat{t} = m_t^2 - (E_t - t_L) x
\surd{s}$ are the differential cross section variables.

{}~~~~Finally, we have to consider the spectrum of the lepton in the top
rest-frame
(and of the antilepton in the antitop rest frame).  With the Standard Model
couplings $W\cdot (\bar{t}\gamma (I+\gamma_5) b)$ and $W\cdot (\bar{l}\gamma
(I+\gamma_5)\nu_l)$, there is a large forward-backward asymmetry in the $W\to
\bar{l}\nu_l$ decay relative to the b-momentum in the W rest-frame. Since there
is a one-to-one relationship between this angle and the energy
$E_{\bar{l}t}$ we
can express this angular distribution as an energy distribution:\\
$$\frac{d \Gamma}{dE_{\bar{l}t}} = G_{F}^{2} M_{W}^{3} E_{\bar{l}t}
(m_t^2 -m_b^2 -2m_t E_{\bar{l}t}) /
(4 \pi^{2} m_t \Gamma_W). \hspace{1.75in} (3.16) $$\\
If we neglect $(m_b/m_t)<< 1$, and normalize the distribution to unity, we have
the probability distribution\\
$$P(E_{\bar{l}t}) = (24/m_t^2) E_{\bar{l}t}(1-2E_{\bar{l}t}/m_t)dE_{\bar{l}t}
\hspace{2.5in} (3.17) $$\\
where $E_{\bar{l}t}$ runs from 0 to $m_t/2$ in this approximation.

{}~~~~Taking all these factors together, the total probability for reaching the
observed configuration is proportional to\\
$$P(\underline{t},\underline{\bar{t}}\mid m_t) = \sum_j \sum_i
F_j(x_i)F_j(\bar{x}_i)\cdot \frac{d\sigma(\hat{s}_i,\hat{t}_i)_j}
{d\hat{t}_i}\cdot
P(E_{\bar{l}t}\mid m_t)P(E_{l \bar{t}}\mid m_t), \hspace{0.75in} (3.18)$$\\
where the sum i is over $(\bar{q}q),(q\bar{q}),(gg)$, and the sum j is over all
of the ellipse-crossing points for the $m_t$ value under consideration.  As
discussed here and in ref.[18], $P(\underline{t},\underline{\bar{t}}\mid m_t)$
is a discontinuous function of $m_t$, the ellipses being lines of zero width
which either cross or don't cross.  In reality, these lines have finite width,
partly because top quarks and W-bosons have finite widths, but more importantly
because of statistical uncertainties in jet development and in measurements of
the particles in the event, and because of a distribution of transverse momenta
for the partons and the antipartons in the incident proton and antiproton.  In
order to estimate $m_t$ from the data on CDF-1, assuming it to be due to top-
antitop production and decay, we appeal to Bayes Theorem, which gives the
probability distribution as:\\
$$P(m_t \mid \underline{t},\underline{\bar{t}}) =
P(\underline{t},\underline{\bar{t}} \mid m_t) \Phi(m_t)/ \{\int dm
P(\underline{t},\underline{\bar{t}}\mid m)\Phi(m)\},  \hspace{0.5in}
(3.19) $$\\
where $\Phi(m)$ is the {\it a priori\/} probability that the top mass is $m$
and
the integral is taken over all possible values of $m$.
With $\Phi(m) = 1$, i.e. for
the CDF-1 event shape alone, the probability $P(m_t \mid
\underline{t},\underline{\bar{t}})$ is plotted in Fig.7(a), as taken from
ref.[18]. Its peak lies at about 130 GeV, and it is broad and flat.

\subsection{\sechead More realistic top mass estimation.}
\vskip -0.05in
\hspace{0.2in}~~~~ Although the above analysis is geometrically simple and
easy
to comprehend, we have come to realise, from the analysis of the
``l+4jets'' events
in Sec.4 below, that it is really necessary to allow (i) for the transverse
momenta of the initial parton-antiparton system, and also (ii) for the
uncertainties of the $b$ and $\bar{b}$ jet energies. In this Section, for
simplicity, we shall consider only the former of these; the inclusion of the
latter is discussed below, in Sec.4.2.

{}~~~~ Consider first the t-ellipse for given $\underline{b}$ and
$\underline{\bar{l}}$, for assumed top mass $m$. Divide its boundary
by N points by
dividing the ellipse of Fig.4 into equal azimuthal slices with respect to the
axis $C_e N$, where $C_e$ denotes the centre of the ellipse and $C_e N$
intersects BL at angle $\sigma$ given by expression (3.11) and is normal to the
plane of the ellipse. At each one of these N points (a set labelled by
$\gamma$)
there is a definite momentum $\underline{t} (\gamma)$. Next, consider the
$\bar{t}$-ellipse for $\underline{b}$ and $\underline{\bar{l}}$,  for the same
value of $m$, following the same procedure
for the N points (this set labelled by $\bar{\gamma}$), each point leading to a
definite momentum $\underline{\bar{t}}(\bar{\gamma})$. We now soften the
condition (3.12) by the introduction of a weighting factor $T_{\rho}$ of finite
range, thus obtaining in place of (3.18):\\
$$P(\underline{t},\underline{\bar{t}} \mid m) = \sum_{\gamma \bar{\gamma}}
\{P(\underline{t}_{\gamma} \mid m) T_{\rho}(\underline{t}_{\gamma T} +
\underline{\bar{t}}_{\bar{\gamma} T}) P(\underline{t}_{\bar{\gamma}} \mid m),
\hspace{1.5in}  (3.20)$$\\
where $T_{\rho}(\tau)$ is the function\\
$$T_{\rho}(\underline{\tau}) = \frac{1}{2\pi
\rho^2}e^{-\frac{\underline{\tau}^2}
{2\rho^2}}, \hspace{2in} (3.21)$$\\
which is a representation of the two-dimensional $\delta$-function
$$T_{\rho}(\underline{\tau})\rightarrow \delta(\underline{\tau}) =
\delta(\underline{t}_{\gamma T} + \underline{\bar{t}}_{\bar{\gamma} T}),
\hspace{2.5in} (3.22)$$\\
as $\rho \rightarrow 0$. In practice, we will use the sum (3.20) over the set
of points
$(\gamma,\bar{\gamma})$, each having 180 points, so that the slices are
of angle
$2^{\circ}$. The contribution of each point to the sum is evaluated and summed.
The largest individual contributions are naturally those from the vicinity
of the
crossing points of Fig.6. The results of this calculation are shown in Fig.7(a)
for $\rho = 0.1 m_t$ and $0.025 m_t$.We see there that the forms obtained
differ
considerably from that obtained with the condition (3.12), and further, that
there is no sign that the earlier result is recovered in the limit $\rho
\rightarrow 0$. However, it is easy to understand the change in form. Consider
the case of $m = 125~GeV$. We no longer have single large contributions
coming only
from the crossing points C and D, but find a large number of smaller
contributions all the way from C to D; this situation persists as the mass
falls below 125~GeV, even after there are no crossings corresponding to C and
D. Since the ellipses become smaller as $m$ falls, the finite range gains
increasing relative importance and it is not surprising that the non-crossing
contributions become dominant there, giving greater weight to low $m$ values
and thus distorting the $m$-distribution away from the form which we presented
in ref.[18].

{}~~~~ The dashed line on Fig.7(a) shows the $m$-probability of the single
point
having the greatest probability for the $m$
considered. This point naturally lies
close to the cross-over point for that $m$, if there is one, so that this
entry must be closely related with the value obtained with the procedure based
on (3.12). This curve, based on these entries of maxima, does have the slow
fall-off for increasingly large $m$ which we noted in ref.[18] and show here in
Fig.7(a). The change introduced by the use of (3.20) arises from the fact that
there are a large number of points $(\gamma,\bar{\gamma})$ which cannot
contribute to the sum $P(\underline{t},\underline{\bar{t}}\mid m)$ when (3.12)
is required but which dominate its sum for low $m$ when (3.20) is used, so
depressing the value of the most likely value for $m$.
The probability distribution
we now adopt is that shown in Fig.7(a) for $\rho = 0.1 m_t$.

\subsection{\sechead  Compatibility and combining of top mass probability
distributions. }
\vskip -0.05in
\hspace{0.2in}~~~~ The probability functions $P(m_t \mid N_{\mu e})$
and $P(m_t\mid \underline{t},\underline{\bar{t}})$ are independent
and at present
compatible. They can be combined to give an overall probability function
from the
data publically available today:\\
$$P(m \mid data) = P(m \mid N_{\mu e}=3) \cdot P(m \mid
\underline{t},\underline{\bar{t}}:CDF-1), \hspace{1.5in} (3.23)$$\\
shown on Fig.7(b). Its peak is at 121~GeV, the one-standard-deviation limits
being 114.5 and 130~GeV. Of course, these remarks are very tentative since
they
are based on the assumption that the three $(\mu^\pm e^\mp)$ events are due to
top production and decay; all three might be due to background.

{}~~~~  At present, we do not know whether or not $m_t = 121(+9,-6.5) GeV$ is
the
top mass. However, the hope for the future is that a substantial number of
$(\mu^\pm e^\mp)$ events will be observed at the Tevatron.
An $m_t$ distribution
can be determined for each of them, as done for CDF-1 here.  These
distributions
may turn out to vary randomly from event to event;  if so, then these events
could not have anything to do with top-antitop production and decay.  More
likely, a large fraction of them (and perhaps all) may peak at a definite
$m_t$
value and we shall be able to conclude that they form a well-defined group of
events due most probably to $t \bar{t}$ production and decay.  The total cross
section for this group of events will provide a quantitative test on this
interpretation for them.

{}~~~~  In order to illustrate one difficulty in following this line of
thought,
we shall close this Section by discussing a fictitious event
which we shall refer
to as $F_{\mu e}$.  Its input momenta are given in Table 4 and it has been
subjected to the same analysis, with $\rho = 0.1 m_t$, as was
applied to the real
event CDF-1. Evaluating the limit (3.8b), the antitop momenta lead to
$(m_t)_{min} = 95.3 GeV$, while the top momenta lead to $(m_t)_{min} = 134.5
GeV$.  Since the latter is the stronger limit, we conclude that, if $F_{\mu e}$
were data from a real $t \bar{t}$ event, then the top mass could not lie lower
than 134.5 GeV.  The $m_t$ distribution for $F_{\mu e}$ is shown on Fig.8.  It
peaks at 154 GeV, with one-standard-error limits at 144 and 168 GeV and it is
almost incompatible with the $m_t$ distribution given in Fig.7(a) for the real
event CDF-1.  We would have to conclude that either one (or both) of
these events
is not the result of top-antitop production and decay, or that there has been a
large statistical fluctuation.  In either case, the situation would be
unsatisfactory and we would have to await further $\mu e$ events.  At
the moment,
as mentioned in Sec. 2(b), it is known that there do exist two further
events of this kind and we all look forward to the
release of the detailed data from them.

\section{\underline{{\sechead ANALYSIS OF ``LEPTON AND FOUR JETS'' EVENTS}}}
\subsection{\sechead Expected rate and nature of the events.}
\vskip -0.05in

\hspace{0.2in}~~~~ As mentioned following eq. (2.9), the hadronic decay modes
$W^{+} \rightarrow u\bar{d}\, and\, c\bar{s}$ have a net rate about
six times greater
than either of the decay modes $W^{+} \rightarrow e^{+}\nu_e \,or\,
\mu^{+}\nu_{\mu}$. From this remark it follows that $t \bar{t}$
decay leading to
the final charged particles\\
$$(l^{+}(=e^{+} \,or\, \mu^{+})b) + \bar{b}(\bar{u}d \,or\,
\bar{c}s), \hspace{2in}
(4.1a)$$\\
and\\
$$(l^{-}(=e^{-} \,or\, \mu^{-})\bar{b}) + b(u\bar{d} \,or\, c\bar{s}),
\hspace{2in}
(4.1b)$$\\
are more frequent than the $(\mu^\pm e^\mp)$ dilepton modes
$$(\mu^{+}b) + (e^{-}\bar{b}), \hspace{3.5in} (4.2a)$$\\
and\\
$$(\mu^{-}\bar{b}) + (e^{+}b), \hspace{3.5in} (4.2b)$$\\
by a factor of twelve.  Here, we have confined attention to the
diagonal elements
of the Cabibbo-Kobayashi-Maskawa matrix and then only for $\lambda = 0$
(Wolfenstein's
parametrization [28]), since this is a satisfactory approximation for the
present. If the three $(\mu^\pm e^\mp)$ events mentioned in Sec. 2(b) were all
due to top-antitop pairs, then we would expect about another 36 pairs, each
giving one energetic lepton with high $p_T$ accompanied by one $b-jet$, one
$\bar{b} -jet$  and two other jets, as specified by the possibilities given in
(4.1).  Up to the present, it has not been possible to identify which are
the $b$
and $\bar{b}$ jets, except by the observation of a low-energy muon
resulting from
the secondary decay $b \rightarrow l^{-} \bar{\nu}_l c$ or $\bar{b} \rightarrow
l^{+} \nu_l $, as was the case in the CDF-1 event, where there was a low-energy
$\mu^{+}$ (see Table 3).  The secondary lepton from $b$ or $\bar{b}$ decay has
branching ratio about $21\%$, so that one should appear in $\approx 42\%$ of $t
\bar{t}$ events; the appearance of two secondary leptons has a rate an order of
magnitude lower. In the present run, CDF is equipped with a
microvertex detector,
which will greatly increase its efficiency for identification of the $b$ and
$\bar{b}$ jets, because of the relatively long lifetime of the b quarks
$(\approx {10}^{-12} sec)$, and so restrict the fitting of the final
state to the
$t \bar{t}$ hypothesis in a very helpful way.
\vskip 0.2in
\begin{center}
{}~Table 4. Input momenta for a fictitious $\bar{t}t$ event ($F_l$)
of the form ($l^{-}b\bar{b}q'\bar{q}$),
where ($q',\bar{q}$) = ($s,\bar{c}$) or
($d,\bar{u}$).
\vskip 0.2in
\begin{tabular}{|c|ccccc|} \hline
GeV/c & $e^{-}$ & jet($\epsilon$) & jet($\psi$) & jet($\phi$) & jet($\eta$) \\
\hline
$p_x$ & 11.0  &  42.0  &   0.7  &    0.0  &    7.0 \\
$p_y$ & -19.0 &  -23.5 &    23.5  &  -25.5  &   16.0 \\
$p_z$ &  -7.0 &   25.0 &   -56.0 &    23.0 &    72.0 \\ \hline
\end{tabular}
\end{center}
\vskip 0.2in

\subsection{\sechead Their analysis and top mass probability distributions. }
\vskip -0.05in

\hspace{0.2in}~~~~Here we discuss the analysis of the $(l + 4jet)$ events, when
there is no information as to which are the $b$ and $\bar{b}$ jets.  What
we say
below will assume the lepton to be the $l^{+}$; if it is $l^{-}$, then
the terms
top and antitop are to be interchanged in what follows.

{}~~~~We start by labelling the jets according to their momenta, using the
symbols
$(\phi ,\psi , \epsilon ~and ~\eta)$, as illustrated in Fig.9 and
in Table 4. The
mass $M_W = 80.2(3) GeV$ and the width $\gamma_W = 2.1(1) GeV$ are now rather
well known for the W bosons;  we shall therefore impose this value of $M_W$ on
the event, while neglecting the width $\gamma_W$ for convenience, since its
effects are of secondary importance.  Two of the jets are selected, say
$\epsilon$ and $\eta$, and identified as 1 and 2, with energies $E_1$ and $E_2$
and separation angle $\theta_{12}$.  We then have\\
$$M_W^2 = m_1^2 + m_2^2 + 2(E_1E_2 - p_1p_2cos\theta_{12})\approx 2E_1E_2(1 -
cos\theta_{12}). \hspace{1in} (4.3)$$\\
Each jet energy $E_i$ has a probability distribution $Q_i(E_i)$, partly because
jet development is a stochastic process and partly because of uncertainties in
its determination from the observations. Eq.(4.3) fixes the product $E_1 E_2$,
so that the allowed values correspond to a hyperbola on the
$(E_1 , E_2)$ plane.
$N_{12}$ points are chosen on that hyperbola and the probability assigned
to each
of these points is deduced from the integral\\
$$Q_{\epsilon \eta}(1,2) = \int dE_1dE_2Q_{\epsilon}(E_1)Q_{\eta}(E_2)
\delta(E_1E_2 -\frac{M_W^2}{2(1-cos\theta_{12})}). \hspace{1in} (4.4)$$\\
A third jet, say $\phi$, is selected and labelled 3; $N_3$ points are
chosen over
the energy range $(E_3 \pm \chi\sigma_3)$, where $\sigma_3$ is the assigned
uncertainty at the one-standard-error level and $\chi$ is chosen suitably
(usually $\chi = 2$).  Thus, we have chosen $N_3 \cdot N_{12}$ energy values (a
set labelled $\alpha$);  for each member of this set, an antitop momentum
$\underline{\bar{t}} (\alpha)$ and mass $m(\alpha)$ are deduced, with
an assigned
probability $Q(\alpha)$, which is the product of $Q_{12}$, computed from (4.4),
and $Q_{\phi} (E_3)$.

{}~~~~  We now turn to $l$ and the fourth jet $\psi$, labelled 4.  This jet is
tacitly (and necessarily) assumed to be a b jet,
although this identification is
not used here.  We take $N_4$ energy values (a set labelled $\beta$) over the
range $E_4 \pm \chi\sigma_4$, each with probability $Q_{\psi} (\beta)$. For
each point $(\alpha ,\beta)$, we determine the ellipse for mass
$m(\alpha)$, based on
the vectors $\underline{\bar{l}}$ and $\underline{b}_{\psi}$, and divide its
boundary by $N_l$ points (a set labelled $\gamma$), as we did in Sec.3.4 by
taking equal azimuthal slices of the ellipse of Fig.4 with respect to the axis
$C_e N$, where $C_e$ denotes the centre of the ellipse and $C_e N$
intersects BL
at the angle $\sigma$  given by expression (3.11) and is normal to the plane of
the ellipse. At each one of these $N_l$ points, there is a definite momentum
$\underline{t}(\alpha ,\beta ,\gamma)$ for the top quark.  Next, we ask for a
match between $\underline{t}$ and $\underline{\bar{t}}$, i.e. that\\
$$\underline{t}(\alpha,\beta,\gamma)_T + \underline{\bar{t}}(\alpha)_T  = 0
\hspace{2.5in} (4.5)$$\\
if the incident parton-antiparton system is required to have no transverse
momentum. However, in general, this equation will not be satisfied, except at
isolated points. As above, in Sec. 3.4, we give up the equality (4.5), allow
initial transverse momentum and introduce the probability distribution (3.21),
where here $\underline{\tau}$ is given by the l.h.s. of eq.(4.5), adopting as
before the value $\rho = 0.1 m_t$. Just as for the dilepton modes in Sec.3, a
lepton factor (3.17) is necessary, which we denote here by $P_{\psi}
(E_{\bar{l}
t}, 4)$. For the accompanying decay sequence $t \rightarrow b W^{-}, W^{-}
\rightarrow \bar{c}s ~or ~\bar{u}d$, the light quark s or d will have the same
strong backward/forward asymmetry as holds for the lepton in $W^{-} \rightarrow
l\bar{\nu}_l$, but since we
have (in general) no means to distinguish the $d$ jet
from the $\bar{u}$ jet (or the $s$ jet from the $\bar{c}$ jet), we have to
average over the two final jets and this removes the
backward/forward asymmetry.
For simplicity, we have taken the decays
$\bar{t} \rightarrow \bar{b}\bar{c}s ~and
{}~\bar{b}\bar{u}d$ to be isotropic.  For the production of
$\bar{t}$ and $t$ quarks
with momenta $\underline{\bar{t}}(\alpha)$ and
$\underline{t}(\alpha,\beta,\gamma)$, through the process $\bar{q}q \rightarrow
\bar{t}t$, there is also a rate factor
$R(x(\alpha,\beta,\gamma),\bar{x}(\alpha))$
necessary, given by (3.15) with $x$ and $\bar{x}$ defined by eqs.(3.14).  Thus,
for the labelling (3,4,1,2) of the four jets $(\phi,\psi,\epsilon,\eta)$ and
for each of the number $(N_l\cdot N_{12}\cdot N_3\cdot N_4)$ of points
defined above,
we have calculated a mass $m(\alpha)$, the two momenta
$\underline{\bar{t}}(\alpha)$ and $\underline{t}(\alpha,\beta,\gamma)$ and an
associated probability for reaching this event configuration, given by the
product\\
$$P(event|m(\alpha);3,4,1,2) = Q_{\epsilon\eta}(1,2)Q_{\phi}(3)Q_{\psi}(4)
T(\alpha,\beta,\gamma)P_{\psi}(E_{\bar{l} t},4)R(x(\alpha,\beta,\gamma),
\bar{x}(\alpha)). ~(4.7)$$\\
Finally, for this particular assignment of the labels (1,2,3,4), the net
probability for this event configuration when $m(\alpha)$ lies in the interval
$(m,m+\delta m)$ is obtained by summing these probabilities (4.7) for all of
these
$(N_l\cdot N_{12}\cdot N_3\cdot N_4)$ points for which their $m(\alpha)$ lies
in this interval.

{}~~~~  When none of the four jets is identified, it is necessary to consider
all
possible identifications for them.  The final expression for the net
probability
that the observed event configuration could occur for top quark mass $m$
is given
by the sum of (4.7) over all permutations of the labels (1,2,3,4):\\
$$P(event\mid m) = \sum_{Perm.} P(event\mid m; 1,2,3,4). \hspace{2in}
(4.8)$$\\
As in Sec. 3 above, we now use Bayes' Theorem to obtain the desired expression
for the probability that the top quark mass is $m$, given the data on this
event
and an {\it a priori\/} probability $\Phi (m)$ from other information (such as
that given by the $\bar{t}t$ production rate observed;  cf.Sec.2.2 above), as
follows\\
$$P(m\mid this ~event + earlier ~info.) = c\cdot P(event\mid m)\Phi(m),
\hspace{1.25in} (4.9)$$\\
where c is a normalization constant, independent of $m$.

{}~~~~  Of course, most of the identifications in the sum (4.8) are necessarily
inappropriate but we have no means, in general, to know which of them
correspond
to the correct physical interpretation.  It is our optimistic expectation that
incorrect identifications will generally lead to negligible contributions to
the
net probability for top mass $m$, although it can certainly happen that several
different interpretations have comparable probabilities.  The merit of this
method of analysing the data on an event is that it is systematic, so that no
possible interpretations can be overlooked, and that it is quantitative,
assigning a numerical value for the relative probability for different
interpretations.  It is not an elegant procedure but it does not involve
multivariate searches for the maximum likelihood.

{}~~~~ We may illustrate the procedure by discussing a
fictitious event of the type
$(l + 4jets)$, with final state momenta as given in Table 4,
which we shall refer
to as $F_l$.  The probability distribution $P(m)$ calculated for $F_l$ is shown
on Fig.8.  This distribution peaks at 135 GeV, with
one-standard-deviation limits
at $\pm 6 GeV$; it has quite a strong overlap with those for CDF-1 and it
overlaps with that for $F_{\mu e}$ no more than that for CDF-1 does.

\subsection{\sechead Simulated data and its analysis. }
\vskip -0.05in

\hspace{0.2in}~~~~To test the method more extensively, we have considered two
sets of simulated data:

\noindent (a) A Toy Model.

{}~~~~This is described in ref.[23].  Adopting a mass of 140 GeV, the vectors
$\underline{\bar{t}}$ and $\underline{t}$ were chosen in a random way, and
their
decay configurations follow the leptonic sequence for t discussed in Sec.3 and
an isotropic 3 jet sequence for $\bar{t}$.  These events were processed by the
procedure described above, including allowance for the Gaussian probabilities
appropriate to the CDF determination of jet energies from the data on each jet
and for all permutations of the assignments of (1,2,3,4) to the jets.  In this
analysis, the same cuts were made on the simulated data as are routinely
applied by CDF to their real data.  These cuts are as follows:

\hspace{0.2in}~~~~(i) $\rho_T > 15 GeV$ for each jet,

\hspace{0.2in}~~~~(ii) $E_{lT} > 20 GeV$ for the lepton,

\hspace{0.2in}~~~~(iii) missing transverse $E_T > 35 GeV$,

\hspace{0.2in}~~~~(iv) pseudorapidity lower than 2.44 in magnitude for all four
jets.

{}~~~~The $P(m\mid event)$ distributions for three "toy events" chosen at
random
from 1000 simulated events are shown in Fig.10.  In two of them, Figs.10(a) and
10(b), the peak of the distribution lies close to 140 GeV, the input value
although with much spreading due to wrong jet combinations.  However the third,
Fig.10(c), shows that individual events can deviate widely from our simple
expectations, for its peak is at 150 GeV while it has a marked dip in the
vicinity of 140 GeV.  In Fig.11, mean probability distributions for 1000 "toy
model" events generated for $m_t = 140~GeV$, are shown for two cases.  For
Fig.11(a), the analysis made allowance for the Gaussian probabilities
appropriate
to the CDF determination of jet energies and excluded events which did not
satisfy the CDF cuts as specified in the last paragraph.  The reader should
note
that this is $\underline{not}$ a distribution of the mean peak mass, but
represents rather the form of a ``typical $m$ distribution''
under the circumstances
specified.  Its peak is at 137 GeV with a width (FWHM) of about 18~GeV.
For Fig.11(b) the jet energies
input were smeared by making random changes in their
magnitudes, with a standard deviation taken from the CDF work, intended to
represent the effects of jet fragmentation and soft-gluon bremsstrahlung and of
the detector efficiencies.  This has depressed the mean probability
distribution
by a factor of about 2/3;  the peak location is  about 1 GeV lower and the FWHM
increased to about 25 GeV.

\noindent (b) ISAJET [24]

{}~~~~This provides a more sophisticated simulation of $\bar{t}t$ production. A
full simulation of the CDF detector effects has been carried out, using the
appropriate jet-finding procedures and fragmentation codes.  These detector
characteristics were taken account of in the code QFL, as understood through
the analysis of the 1988-89 CDF run [24].  The standard ISAJET program does not
take account of t decay to $bW^{+}$, the process which dominates over t-jet
fragmentation for the top mass values of interest to us here;  however, ISAJET
does have the option of adding one additional channel for transitions of the
top quark, beyond those for the usual jet development, and this has been made
use of here. This channel then becomes the dominant one, so that the later jet
development is entirely that of the secondary b-quark from top decay and of the
tertiary ($c$ and $\bar{s}$ or $u$ and $\bar{d}$) quark decays.  The 500
$\bar{t}t$ production and decay events generated in this way for $m_t = 140
{}~GeV$ have been processed as discussed above for systems with the
$(l^{+}+4jets)$ final state structure and the resulting mean probability
distribution has been plotted in Fig.12. This distribution does show a quite
definite peak at 130 GeV, a shift of 10 GeV below the input mass, with a FWHM
of about 25 GeV if the
secondary peak
at about 108 GeV is ignored. This relatively large shift is believed to be due
the gluon radiation which is taken into account in ISAJET, but this is still
under investigation. The secondary peak is believed to arise from the ``wrong''
combinations of jets but this is not yet certain; however, it appears unlikely
to interfere with the use of the upper peak to determine the top quark mass,
since it may be masked by the low-mass contributions to $P(m\mid event)$ from
background events. We note also that the mass distributions predicted for the
ISAJET model spread to much higher top-mass values than do the simpler ``toy
model'' calculations.

{}~~~~ These model calculations demonstrate that when sets of input data
obtained
from ``events'' generated by an algorithm representing any of the processes\\
$$\bar{p} + p \rightarrow \lbrace{{e^{+}+\mu^{-}} \atop {e^{-}+\mu^{+}}}\rbrace
{}~b +\bar{b} + \lbrace{{\nu_e + \bar{\nu}_{\mu}}\atop{\bar{\nu}_e +
\nu_{\mu}}}
\rbrace + X, \hspace{1.2in} (4.10)$$\\
or their charge conjugates, which have proceeded through the hadronic
$\bar{t}t$
production process (2.4),
are considered, the method of analysis we have proposed
[23] does lead to $m$ distributions which peak
strongly at a mass close to the top
mass used in generating these events, at least on the average. This holds true
even when we do not know the identity of the quark or gluon which has generated
each of the observed jets. However, we must next enquire whether a systematic
peaking of the $m$ distributions computed from real data at some particular
mass
value, say m*, necessarily implies the existence of a top quark with a
mass close
to m*. Are there other processes leading to the final states $(l+4jets)$, which
we might describe as ``background'', for which our analysis leads to such peaks
even though there may be no top quark at all or perhaps just no top quark in
the mass range explored? We shall see that the answer is "yes" because the
distribution P(m) from the background must vanish at the threshold $(M_W+m_b)$
and approach zero again as m becomes sufficiently large since the parton
parameters $\alpha$ and $\bar{\alpha}$ cannot exceed unity. Precisely where its
maximum occurs depends on the nature of the background and on what cuts are
made
to reduce its contribution to P(m) in the mass region accessible for the top
quark.

\subsection{\sechead Simulation of non-top background and its analysis as
if top-antitop }
\vskip -0.05in

\hspace{0.2in}~~~~It is apparent that final states of the form\\
$$\bar{p} + p \rightarrow l  + 4 jets + \nu_{l} + X \hspace{2in} (4.11)$$\\
can readily arise without top production, through single W production and
decay, as follows:\\
$$\bar{p} + p \rightarrow W + (4 ~hadronic ~jets = \bar{q}\bar{q}qq ~or
{}~\bar{q}qgg
{}~or ~gggg), ~W\rightarrow l + \nu_l. \hspace{0.25in} (4.12)$$\\
Berends,{\it et al.\/}[26]
have provided full tree-level calculations of the cross
sections for all of the Standard Model processes which involve one W-boson and
$n \leq 6$ quarks, antiquarks and gluons. The case $n=6$ includes all of these
processes (4.11) and therefore provides a natural and plausible $"l + 4 jets"$
background to the $"l + 4 jets"$ states (4.1) which result from the sequence of
$\bar{t}t$ production (2.3), their decays (2.6) and finally leptonic decay for
one W and decay to three quarks for the other, as outlined in (4.1). In the
latter sequence, all four jets are quark or antiquark jets, one being a $b$ jet
and another a $\bar{b}$ jet. The calculation by Berends,{\it et al.\/}
for the processes
(4.12) at the Tevatron energy predicts, after application of the cuts specified
in Sec. 4.3 above, that the set of (one quark, one antiquark and two
gluon)-jets
occur more often $(56\%)$ than the set of (two quarks and two antiquarks)-jets
$(42\%)$, while the set of four gluon jets is quite rare $(2\%)$. It predicts
also that, with these cuts, $b$ and $\bar{b}$ jets should occur among
these quark
jets only rarely $(3\%)$. Berends, {\it et al.\/} also calculated
the rate for the
$\bar{t}t$ production process (2.4) in the same approximation, with the decay
of the consequent $W^{+}$ or $W^{-}$ boson leading to the final states (4.10).
Applying the same cuts to these final states, they then concluded from their
calculation that, for top masses between 100 and 135 GeV, the number of these
events due to $\bar{t}t$ production and decay would exceed the number of
background events.

{}~~~~ The $m$ distributions obtained by subjecting these QCD background events
to our analysis procedure may be expected to differ appreciably from
those for real
$\bar{t}t$ production and decay. We have examined this question by using the
VECBOS Monte Carlo program[27] which implements these Standard
Model calculations
of Berends, {\it et al.\/} for "W+3jets" and "W+4jets" events, to obtain a
large
sample of calculated events which are input to the ISAJET+QFL evolution and
development. The sample of "W+3jets" used corresponds to an $IL = 112~pb^{-1}$
and that of "W+4jets" to $IL = 128~pb^{-1}$. Each event was put through our
analysis procedure as described above, assuming it to result from $\bar{t}t$
production and decay, and gave an $m$
probability distribution $P(m\mid l+4jet)$.
The mean $m$ distribution, normalised to $IL = 4~pb^{-1}$,
is shown by the shaded
area on Fig.13. It does show a peak at $m\approx 112~GeV$, with an appreciable
tail running up as high as 140~GeV. The open area of Fig.13 corresponds to the
fictitious event $F_l$ shown on Fig.8, with its peak at 135~GeV; the two peaks
are quite well separated, showing that a $\bar{t}t$ event of type $F_l$ with
top
mass 135~GeV could readily be separated from this background, given sufficient
events. However, if the top mass were much lower than 120~GeV - the value our
analysis of the data on CDF-1 might suggest - the $"l+4jets"$ $m$ distribution
would overlap so strongly with the background $m$ distribution as to make their
separation more problematical, requiring a very
large body of data, at the least.
In this situation,the most convincing evidence on $m_t$ would be that
obtainable
from the $(\mu^\pm e^\mp)$ dilepton events,
where no source of serious background
is known.

\subsection{\sechead $b-$ and/or $\bar{b}-$quark tagging. }
\vskip -0.05in

\hspace{0.2in}~~~~This situation could be much improved if we had some
knowledge
of the identities of the four jets in these $"l+4jets"$ events, since some
permutations of the labels (1,2,3,4) would then be excluded and the spread
of the
$P(m_t \mid l+4jets)$ distribution due to incorrect identifications
would be much
reduced. For example, for our fictitious event $F_{\mu e}$, if jet($\phi$) had
an accompanying low-energy lepton $l^{+}$ in roughly the same direction, this
would suggest that jet($\phi$) in Table 4 is the $\bar{b}$ jet, since the two
leptons $e^{-}$ and $l^{+}$ would then stem naturally from the $\bar{t}$ decay
sequence $\bar{t} \rightarrow \bar{b}W^{-}$, followed by
$\bar{b} \rightarrow \bar{c}l^{+}\nu_l$ and $W^{-} \rightarrow
e^{-}\bar{\nu}_e$.
In this case, jet($\phi$) would be assigned the label 4 and
the only permutations
which need be made are those within (1,2,3). On the other hand, if the
secondary lepton were an $l^{-}$, then jet($\phi$) would be the b jet and
should be assigned the label 3 in the above discussion; only the permutations
within (1,2,4) would be necessary in the sum for the net probability. In
Sec.4.1, we noted that the emission of a lepton from either the $b$ or the
$\bar{b}$ quark will occur in about $42\%$  of the $\bar{t}t$ events (2.5), so
that observation of one secondary lepton should be a common occurrence.
Sometimes this lepton may escape detection because of its relatively low
energy, so that, in practice, its rate may not be as high as this remark would
suggest, but whenever it can be observed and measured, it can be made use of.

{}~~~~ Now that microvertex detectors are being used by CDF, it will often be
possible to identify the $b$ and $\bar{b}$ jets from the finite visible (or
inferred) path up to the decay of the $B^{\pm}$ and $B^{0}(\bar{B}^{0})$
mesons
to which they give rise. With a magnetic field present, measurement of
the charge
signs of the B decay products will enable $B^{+}$ and $B^{-}$ to be separated.
The existence of $(B_s^0,\bar{B}_s^0)$ mixing, with quite a high rate, will
complicate their assignment to $b$ or $\bar{b}$ jet. However, the formalism to
take this mixing into account is well-known and without free parameters (apart
from those concerning CP violation, which makes no appreciable contribution to
the hadronic phenomena under discussion here), so that it can be built into the
calculation of $P(m\mid l+4j)$ from the data. Observations with the microvertex
detectors should both support and supplement the observations of secondary
leptons, but even together they will not make the identification of the $b$ jet
or $\bar{b}$ jet possible in all events. Nevertheless, such "b-tagging",
whether from secondary leptons or from determining the flight time from source
to decay for the b quark, will prove to be of great value both for reducing the
number of misinterpreted background events and for identifying some of the
quark and antiquark jets, thus reducing the number of irrelevant jet
assignments included in the determination of $P(m\mid l+4j)$ from the data.

\section{\underline {{\sechead DISCUSSION}}}
\vskip -0.05in
\hspace{0.2in}~~~~It is clear that we can do little more at present beyond
discussing possible procedures for the analysis of top-antitop candidate
events.
There is little data and it suffers serious uncertainties, so that we are not
able even to test our theoretical assumptions yet. Nevertheless, it is natural
to assume that the top quark exists with some mass value yet to be determined,
and that it does decay dominantly through $t \rightarrow bW^{+}$, as
the Standard
Model indicates for the mass values we must consider today. We believe that we
should launch into the discussion of all candidate events. Even if it is not
possible to prove that individual events are necessarily top-antitop, we may
still find that there is a substantial fraction of the $(\mu^\pm e^\mp 2jets)$
events that give $P(m\mid event)$ distributions which are compatible and can be
combined, in a first step to the determination of the top mass. We have
illustrated this here by combining the $P(m\mid CDF-1)$ distribution with the
$P(m\mid N_{\mu e})$ determined from the rate of observed top-antitop pair
production at the Tevatron; they are compatible and combining them gives an
improved estimate of $m_t$.

{}~~~~We have noted above that $"l+4jets"$ events from the decay of a
$\bar{t}t$
pair are expected to occur twelve times as often as $"\mu^\pm e^\mp 2jets"$. If
the three events of the latter type are really due to $\bar{t}t$, there should
also be, in the recent data, roughly $12\times3 = 36$ events of the former
type.
At present, there do not appear to be even as many candidate events as this
remark suggests, whereas the majority of such candidate events are
not compatible
with top mass as high as the analysis of CDF-1 and the $(\mu^\pm e^\mp)$ event
rate would suggest. This channel calls for open discussion of all $"l+4jets"$
events. It could prove to be a very fruitful line of investigation, where the
advantage of much higher statistics has to be balanced against the necessity
for discriminating against background events. It is worth remarking here that
candidate events are a most powerful stimulus to a theoretician's thinking; it
is not good for the progress of particle physics that experimenters should hold
the details of new events unnecessarily long.

{}~~~~ Berends {\it et al.\/} have provided an excellent QCD model for non-top
production of $"l+4jets"$ events, whose implications must continue to be
studied
further. Nevertheless, this is only a tree-level model and the proposal
and study
of other, no doubt more elaborate, models of background will merit much more
attention. However, these models will not come forth in the absence of a
knowledge of the nature of the events which are actually occurring. For
example, there are certainly events which have more than the minimal number of
jets, such as $"l+njets"$ coming from $\bar{t}t$ production as well as from its
background of W-production and decay, where $n\geq 5$. We shall need
quantitative estimates of the rate of these more complex final states, which
compete with the simpler states having the minimal number of jets and which so
reduce the rate of the latter.

{}~~~~ The observation of the rate $N_{\mu e}$ of the production of the states
$(\mu^\pm e^\mp)$ for high energy leptons is a direct measure of $\bar{t}t$
production. The competing source of $(\mu^\pm e^\mp)$ is $W^{+}W^{-}$ pair
production, as pointed out in Sec.2.2, which is anticipated to have
a lower rate.
Both of these processes, $(\bar{t}t+jets)$ and $(W^{+}W^{-}+jets)$ are also
calculable at tree level. The former modifies the cross section for $(\mu^\pm
e^\mp)$, while the latter provides background. The determination of $m_t$ from
the rate of $(\mu^\pm e^\mp)$ events also needs attention to theory beyond
tree
level. The simplest calculations indicate that the $N_{\mu e}$ rate at energy
$2E_p$ falls rapidly with increasing $m_t$, at the Tevatron energy, and it is
reasonable to expect that the corrections just mentioned will not change this.
The conclusion is that the value of $m_t$ obtained from $N_{\mu e}$ should be
relatively stable with respect to experimental errors in $N_{\mu e}$,
such as the
inclusion of uncertain background events, as we noted in Sec.2.2.

\section{\underline {{\sechead CONCLUSION}}}
\vskip -0.05in
\hspace{0.2in}~~~~No firm conclusion can be expected from this work so far,
except that we should be active in analysing all of the candidate events as
they
emerge, and that the $"\mu^\pm e^\mp2jets"$ events have a special value since
they suffer little background from non-top events. It will be possible
to combine
the $P(m_t)$ probabilities from different events and event types as the data
improves. One advantage of the relatively low energy of the Tevatron is that
two
$\bar{t}t$ pairs will be produced in the same event only very rarely, so that
this source of confusion will be essentially absent. Of course, a
somewhat higher
energy would have the advantage of increasing single $\bar{t}t$
production beyond
the low rate possible from the present Tevatron. The proposed increase of its
c.m. energy from 1800 to 2000 GeV this Summer, with other upgrades, will raise
the rate by a factor of two; clearly an optimum energy would lie still somewhat
higher than this. We shall not try here to estimate what would be the optimum
energy, by balancing a high rate of single $\bar{t}t$ pairs against a
higher rate
of multiple $\bar{t}t$ pairs, since it would be a rather academic exercise
at the
moment. However, it is a question worthy of serious consideration to give
a well-judged answer.

{}~~~~Finally, we must emphasize that no top candidate yet reported has been
demonstrated to represent $\bar{t}t$ production and decay, and that it may
still be some considerable time before we reach that stage of certainty.
However, we understand that the present plans at Fermilab are to run on until
the end of May 1993, by which time the integrated luminosity IL should reach
about $50 pb^{-1}$. The upgrade to 2000 GeV will take place in Summer 1993, and
Tevatron running should begin again in the following October and run steadily
all through 1994. Thus, it is hoped that the net IL reached by the end of 1994
may be about $200 pb^{-1}$, a very substantial improvement over the IL up to
the end of 1992, with which this paper has been concerned.

\section{\underline {{\sechead ACKNOWLEDGEMENTS}}}
\vskip -0.05in
\hspace{0.2in}~~~~The first author (R.H.D.) thanks Prof.A.Zichichi for the
opportunity to speak in this Summer School at ERICE, the site of so many
important physics meetings, and Profs.D.Sherrington and G.G.Ross for the
continuing hospitality of the Theoretical Physics Department of the University
of Oxford, and the second author (G.R.G.) wishes to acknowledge the
U.S.Department of Energy for partial support during this work. We have
appreciated discussions with Dr.K.Sliwa and the help of Drs.J.Benlloch, M.Timko
and N.Wainer of C.D.F. with various aspects of the simulations described
briefly here.

\section{\underline {{\sechead REFERENCES}}}
\vskip -0.05in

 1. G.L.Kane, Top Quark Physics, in Proc. Workshop on High Energy
 Phenomenology at Mexico City,1-10 July,1991. Also issued as Univ. of
    Michigan preprint UM-TH-91-32 (December,1991).

 2. CDF Collaboration,F.Abe et al.,Phys.Rev.Lett.68(1992);
ibid, Phys.Rev.D45 (1992)3921.

 3. C.A.Nelson, Phys.Rev.D41 (1990)2805.

 4. G.L.Kane,G.A.Ladinsky and C.-P. Yuan, Phys.Rev.D45 (1992)124.

 5. R.H.Dalitz and Gary R.Goldstein,Phys.Rev.D45 (1992)1531.

 6. I.Bigi,Phys.Lett.B175 (1986)233; I.Bigi et al., ibid, 181(1986)157.

 7. F.J.Gilman and R.Kauffman, Phys.Rev.D37 (1988)2676.

 8. R.H.Dalitz,Gary R.Goldstein and R.Marshall,Z.Phys.C 42(1988)441.

 9. M.Veltman,Nucl.Phys.B 123(1977)89.

10. J.Ellis and G.L.Fogli,Phys.Lett.B 232(1989);
ibid, 249 (1990)543; J.Ellis,
    G.L.Fogli and E.Lisi, Phys.Lett.B 274(1992)456.

11. F.del Aguila,M.Martinez and M.Quiros,preprint CERN-TH.6389/92.

12. U.Amaldi,W.de Boer and H.Furstenau,Phys.Lett.B 260(1991)447.

13. U.Amaldi et al.,Phys.Rev.D 36(1987)1385.

14. J.Ellis,G.L.Fogli and E.Lisi, Phys.Lett.B 287(1992)335; ibid, 292(1992)427.

15. E.Eichten,in Physics at Fermilab in the 1990's. World Scientific,
    Singapore (1990)p.56.

16. D.Crane, in Physics in Collision XI (ed. J.-M.Brom, D.Huss
and M.-E.Michalon, Editions Frontieres,1991)p.145.

17. CDF Collaboration, F.Abe et al.Phys.Rev.Lett. 64(1990)147;
K.Sliwa, in  Proc.XXVth Intl.Symp.on Heavy Flavors Physics, Orsay,France,
1991(eds. M.Davier and G.Wormser, Editions Frontieres,
Gif-sur-Yvette,1992)p.567.

18. R.H.Dalitz and Gary R.Goldstein,Phys.Lett.B 287(1992)225.

19. K.Kondo,J.Phys.Soc.Japan 57(1988)4126; ibid,60(1991)836.

20. R.P.Feynman,Phys.Rev.Lett.23(1969)1415.

21. R.G.Roberts, The Structure of the Proton. (C.U.P.,Cambridge, England,
 1990).

22. D.W.Duke and J.F.Owens,Phys.Rev.D 30(1984)49.

23. Gary R.Goldstein, K.Sliwa and R.H.Dalitz, Phys.Rev.D 47(1993)967.

24. F.Paige and S.Protopopescu,in Physics of the Superconducting Super
 Collider, Snowmass,1987, Proc.Summer Study, Snowmass, Colorado, 1986 (ed.R.
    Donaldson and J.Marx, Division of Particles and Fields of the American
    Physical Society, New York, 1986)p.320.

25. C.Newman-Holmes and J.Freeman, in Proc.Workshop on Detector Simulation for
    the SSC, Argonne Natl.Lab., 1988)pp.190 and 285. Also issued as Rept.No.
    ANL-HEP-CP-88-51.

26. F.A.Berends, H.Kuijf, B.Tausk and W.T.Giele,Nucl.Phys.B 357(1991)32.

27. We are indebted to Jose Benlloch for generating the VECBOS samples used.
The cuts imposed at the generator level differed from the cuts used for the Toy
Model and the ISAJET calculations given here in Sec.4(a). Jets were defined by
the condition  $R =\surd(\delta\phi)^2 + (\delta\eta)^2) >0.6$, where $\phi$
denotes the azimuth angle about the direction of the jet and $\eta$ denotes the
pseudo-rapidity, with $p_T > 10 GeV$ and $|\eta| < 2.5$; the corresponding
lepton cuts were $p_{lT} >10 GeV$ and $|\eta| < 1.2$.

28. L.Wolfenstein, Phys.Rev.Lett.51(1983)1945.

29. M.Jezabek and J.H.K\"uhn, "The top width: theoretical update'', Karlsruhe
preprint TTP 93-4 (1993).

\section{\underline {{\sechead FIGURE CAPTIONS}}}
\vskip -0.05in
\hspace{0.2in}~~~~

1. The partial lifetime given by the Standard Model for $t \rightarrow
be^{+}\nu_e$ is plotted vs. top mass $m_t$. The factor 9 ensures that for $m_t
> 100 GeV$ the quantity  plotted is the total top quark lifetime.

2. The mean number of $(\mu^\pm e^\mp)$ events expected for
integrated luminosity
$30 pb^{-1}$ for $\bar{p}p$ collisions at total energy 1800 GeV, as calculated
by Eichten[15] and Crane[16].

3. Shows the probability distribution for $m_t$, given that there are
3 $\mu^\pm e^\mp$ events observed for integrated luminosity $30 pb^{-1}$.

4. Shows the ellipse PQ comprising all configurations for the vectors
$\underline{t}$,$\underline{b}$ and $\underline{l}$ when the top mass is $m_t$.

5. The $t$ and $\bar{t}$ ellipses for the event CDF-1, as function of $m_t$,
specified at 10 GeV intervals.

6. Projections of the $t$ and $\bar{t}$ ellipses for the CDF-1 event onto the
plane perpendicular to the beam direction for a number of $m_t$ values.

7. Shows (a) $P(m_t\mid CDF-1)$ for zero initial transverse momentum and for
Gaussian initial transverse momentum with s.d. $\rho = 0.025 m_t$ and
$0.1 m_t$;
for the dashed curve, see text. (b) $P(m_t\mid N_{\mu e}=3 ~and ~CDF-1)$,
for $\rho = 0.1 m_t$.

8. Shows $P(m_t\mid F)$ for the fictitious events $F_{\mu e}$ and $F_l$,
described in the text.

9. The leptons and jets in a typical $"l+4jets"$
configuration (actually for the
fictitious event $F_l$), following $\bar{t}t$ production and decay. A low-
energy secondary muon, from $b$ or $\bar{b}$ decay, is added, as indicated by
a short dashed line.

10.Three randomly chosen $"l+4jets"$ events from the Toy Model (Sec.4.3).

11.The mean probability distribution obtained (a) from 1000 Toy Model events
 and (b) from the same events after "smearing" (see text above, also
ref.[23]) to simulate real events, random changes being made in the jet
energy magnitudes to represent effects of jet fragmentation, soft gluon
bremsstrahlung, and detector efficiencies.

12.The mean $m_t$ distribution from our analysis of 500 production and decay
   events generated by the ISAJET+QFL programs, taking into account fully
   the CDF detector characteristics.

13.The mean $m_t$ distribution from our analysis of Standard Model $"l+4jets"$
events generated by VECBOS simulating the tree-level calculations of
backgrounds events by Berends {\it et al.\/}[26]; cuts as in Sec.4.2. The
sample was about 60 times that appropriate for the 1988-89 CDF run.

\end{document}